%% file: main.tex
\documentclass{article} %

\PassOptionsToPackage{numbers,sort&compress}{natbib}

\usepackage[preprint]{neurips_2023}

\input{packages}

\title{RecFusion: A Binomial Diffusion Process for 1D Data for Recommendation}

\author{Gabriel Bénédict\\ %
        University of Amsterdam and RTL Netherlands\\
        \texttt{g.benedict@uva.nl} \\
        \And
        Olivier Jeunen\\
        Sharechat\\
        \texttt{jeunen@sharechat.co}
        \And
        Samuele Papa\\
        University of Amsterdam \\
        \texttt{s.papa@uva.nl} \\
        \And
        Samarth Bhargav\\
        University of Amsterdam \\
        \texttt{s.bhargav@uva.nl} \\ 
        \And
        Daan Odijk\\
        RTL Netherlands \\
        \texttt{daan.odijk@rtl.nl} \\ 
        \And
        Maarten de Rijke\\
        University of Amsterdam \\
        \texttt{m.derijke@uva.nl}
        }

\begin{document}

\maketitle

\begin{abstract}
In this paper we propose RecFusion, which comprise a set of diffusion models for recommendation. Unlike image data which contain spatial correlations, a user-item interaction matrix, commonly utilized in recommendation, lacks
spatial relationships between users and items. We formulate diffusion on a 1D vector and propose \textit{binomial diffusion}, which explicitly models binary user-item interactions with a Bernoulli process. We show that RecFusion approaches the performance of complex VAE baselines on the core recommendation setting (top-n recommendation for binary non-sequential feedback) and the most common datasets (MovieLens and Netflix). Our proposed diffusion models that are specialized for 1D and/or binary setups have implications beyond recommendation systems, such as in the medical domain with MRI and CT scans.
\end{abstract}

\input{sections/01_introduction.tex}

\input{sections/02_method.tex}
\input{sections/03_experimental_setup.tex}
\input{sections/04_results.tex}

\input{sections/05_related_work.tex}

\input{sections/06_discussion.tex}

\newpage

\bibliographystyle{plainnat}
\bibliography{refs}

\input{appendix.tex}

\end{document}

%% file: packages.tex
\usepackage[utf8]{inputenc} %
\usepackage[T1]{fontenc}    %

\usepackage{amsmath}
\usepackage{amssymb}       %
\usepackage{bm}
\usepackage[dvipsnames]{xcolor}

\usepackage[inline]{enumitem}
\usepackage{paralist}

\usepackage{hyperref}       %
\usepackage{url}            %
\usepackage{booktabs}       %
\usepackage{amsfonts}       %
\usepackage{nicefrac}       %
\usepackage{microtype}      %
\usepackage{cleveref}       %
\usepackage{lipsum}         %
\usepackage{graphicx}

\usepackage[inline]{enumitem} %

\usepackage[skip=0pt]{caption}
\usepackage{subcaption}
\usepackage{multirow}

\usepackage{wrapfig}

\usepackage{adjustbox}

\usepackage[resetlabels]{multibib}

\newcites{web}{Additional References}

%% file: sections/01_introduction.tex
\section{Introduction}
Diffusion models have been profusely used in the image domain~\citep{croitoru-2023-diffusion}. Next to the 2D setup, an increasing amount of research is focused on the 3D domain \citep{ho-2022-video}, as well as diffusion on the embedding space \citep{gao-2023-difformer}.\footnote{This line of work on the embedding space is still emergent, we could only find a work in the text domain.} Typical image diffusion models rely on a U-Net \citep{ronneberger-2015-unet} architecture with attention layers and process entire images at once. However, image diffusion models rely on and exploit spatial correlations (i.e., between pixels in localized regions). Unstructured data settings, where these assumptions do not hold, are under-explored. %
In this paper, we consider the recommendation systems domain and, more specifically, 1D binary data in the classical recommendation setting. The recommendation setting is characterized by the following conditions:
\begin{enumerate*}[label=(\roman*)]
\item a user's interaction history is organized like 1D binary data, where columns represent items; and
\item organized as a matrix for multiple users, each entry in this \textit{interaction matrix} corresponds to the type of interaction between a specific user and item. These interactions
can either be \textit{explicit} (ratings, `likes' or dislikes), or \textit{implicit} (dwell time, clicks, purchases, etc.). 
\end{enumerate*}

Most modern recommender systems leverage the implicit feedback paradigm, which utilizes data that is not explicitly provided by the user, such as click data, purchase history, browsing behavior. %
Research in recommender systems employs simpler linear models \cite{nonNeuralClassic2, nonNeuralClassic3, nonNeuralClassic4, nonNeuralSparse}, or neural models, many of which employ the variational autoencoder \cite{kingma2013:vae} framework, e.g., cVAE~\citep{chen-2018-collective}, RecVAE \cite{RecVAE} or MultVAE \cite{liang2018:multvae}. 
Neural models have benefits beyond recommendation performance (e.g., in controllability / critiquing \cite{gao2019:dlc, li2020:rankopt, luo2020:deepcritiquing, yang2021:bayesiancrit}), with some models utilizing disentanglement \cite{bengio2013:repr_learning, higgins2017:betavae} for this purpose \cite{ma2019:macridvae, nema2021:betavae, bhargav2021:controlrecsys}. 
Beyond controllability, neither non-neural nor VAE-based models can handle time information directly. Making it hard, for example, to deal with preference drift~\citep{huang-2022-different}, where more recent items may be more relevant for future recommendations.
In principle, diffusion models should be able to deal with these recommendation conditions.
There have been some initial attempts at modeling recommendation problems using diffusion; CODIGEM~\citep{walker-2022-recommendation} defines a diffusion model akin to early attempts in diffusion, with one neural network per diffusion step. We propose RecFusion, a diffusion model inspired by the DDPM~\cite{DDPM} architecture adapted for 1D data. We also propose the Bernouilli diffusion process, specifically designed for binary data. We experiment with different common diffusion techniques, such as noise timestep embeddings, modelling the mean and variance and different noise schedules.

\begin{figure*}[t]
    \centering
    \includegraphics[width=\textwidth]{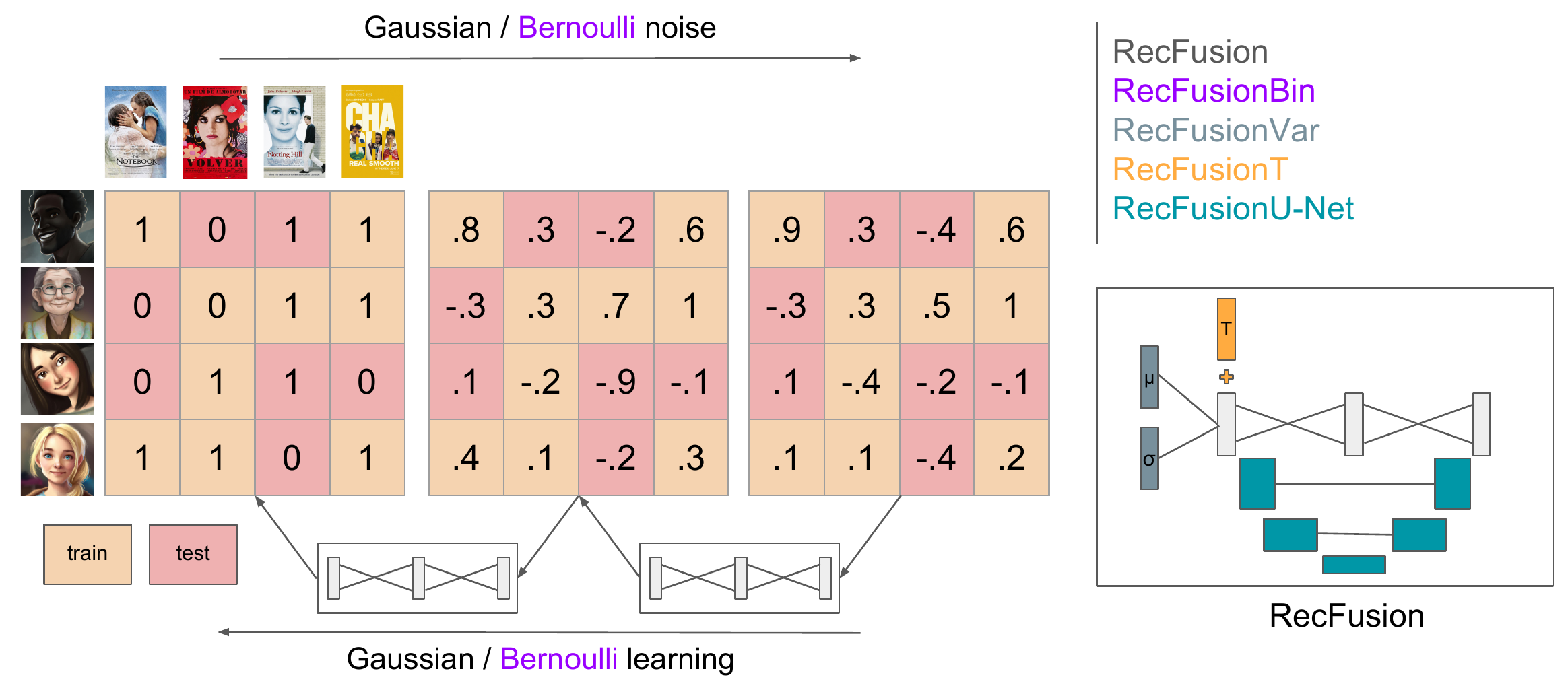}
    \vspace{.3\baselineskip}
    \caption{The RecFusion architecture and its variations (user images generated with DALL·E 2).}
    \label{fig:one}
\end{figure*}

\textbf{Setup.} We assume a binary non-sequential top-n implicit feedback setting (see explicit assumptions in Section~\ref{sec:method}): we seek to predict only the immediate next best item(s) for each user and the time is unknown for any past user-item interaction. The reason for choosing this standard setup is two-fold:
\begin{enumerate*}[label=(\roman*)]
\item by using binary data, we can study the use case of diffusion for binary data, and 
\item we remain comparable with the overwhelming majority of recommendation literature. Indeed, it is common for Assumptions 1--5 that we specify in Section~\ref{subsec:assumptions} to be used.
\end{enumerate*}

\textbf{Main results.} As previously shown in the literature, VAE-based models and non-neural models outperform more complex methods in the standard recommendation setting. RecFusion outperforms existing diffusion models for recommendation and opens the way to use guidance and conditioning.

\textbf{Key contributions} of the paper include: 
\begin{enumerate*}[label=(\roman*)]
\item we demonstrate uses of diffusion 
where there is no spatial dependency, 
\item we offer a simple implementation of diffusion that can accomodate any binary and/or 1D data,
\item we propose modern Variational Auto Encoder (VAE) architectures for  recommendation (diffusion models are hierarchical VAEs~\cite{VDM})
\item our code is open and available at \url{https://github.com/gabriben/recfusion}, implemented using a reproducible and well-tested framework to facilitate follow-up work.
\end{enumerate*}

%% file: sections/02_method.tex
\section{Method}
\label{sec:method}

\subsection{Diffusion models}
\label{subsec:diff}

Diffusion models \cite{diffusion} are latent variable models that strive to address the issues of tractability and flexibility by gradually converting a distribution into another using a Markov chain. The gradual change from the intractable data distribution to a known one allows for the reverse process to be learned \cite{Feller1949OnTT}. In simple terms, this solves traceability, as now one can obtain samples of the data distribution by starting from the known one and using the learned reverse process. At the same time, it introduces flexibility, thanks to the compounding effect of many small simple steps, which allow for the complexity of the target distribution.

Given the starting binary variable $\mathbf{X}^0$, the forward diffusion process is a Markov chain used to sample latent variables $\mathbf{X}^1, \dots, \mathbf{X}^t, \dots, \mathbf{X}^T$. We try to match the original notation from \cite{diffusion}, and make a few simplifying assumptions that are clear from context. In order to generalize, we use the notation $\mathbf{X}$ to indicate a 2D user-item matrix composed of user vectors $\mathbf{x}_u$,\footnote{In most neural recommendation algorithms, a single vector $\mathbf{x}_u$ is fed to the model, see Section~\ref{subsec:recdiff} for a discussion on that topic.} themselves composed of individual interactions $x_{u, i}$. %
The \textit{forward diffusion process} gradually adds noise using a Markov kernel $\kappa_p(\mathbf{X}|\mathbf{X}';\beta)$, until the target distribution $p(\mathbf{X})$ is reached using diffusion rate $\beta$. The kernel used is constructed such that it guarantees that the target distribution is reached in the limit of $T \rightarrow \infty$.

The forward diffusion process is factorized as follows:
\begin{align}
q\left(\mathbf{X}^{1: T} \mid \mathbf{X}^{0}\right) &=\prod_{t=1}^{T} q\left(\mathbf{X}^{T} \mid \mathbf{X}^{t-1}\right)\\
q\left(\mathbf{X}^{T} \mid \mathbf{X}^{t-1}\right) &= \kappa_{p}\left(\mathbf{X}^t|\mathbf{X}^{t-1}; \beta_t\right).
\end{align}
In this work, the kernel $\kappa_{p}$ is either given by the original gaussian diffusion~\cite{diffusion} or binomial diffusion (see next section).
Then, we can define the \textit{reverse diffusion process} using an analogous formulation:
\begin{align}
p_{\theta}\left(\mathbf{X}^{0: T}\right)&=p\left(\mathbf{X}^{T}\right) \prod_{t=1}^{T} p_{\theta}\left(\mathbf{X}^{t-1} \mid \mathbf{X}^{t}\right).
\end{align}
Thanks to the use of very small diffusion steps, the functional form of the reverse process is the same as the forward process \cite{Feller1949OnTT}.

The noise schedule of the forward diffusion process $\beta_{1}, \ldots, \beta_{T}$ is either learned or follows a predetermined schedule (increasing, decreasing or constant). 
The optimization of the backwards diffusion process follows the classical Evidence Lower Bound (ELBO) formulation:
\begin{align}
\mathbb{E}_{q}[\underbrace{D_{\mathrm{KL}}\left(q\left(\mathbf{X}^{T} \mid \mathbf{X}^{0}\right) \| p\left(\mathbf{X}^{T}\right)\right)}_{L_{T}}+\sum_{t>1} \underbrace{D_{\mathrm{KL}}\left(q\left(\mathbf{X}^{t-1} \mid \mathbf{X}^{T}, \mathbf{X}^{0}\right) \| p_{\theta}\left(\mathbf{X}^{t-1} \mid \mathbf{X}^{T}\right)\right)}_{L_{t-1}} \hspace*{5mm}\mbox{}\nonumber \\[-2mm]
{}-\underbrace{\log p_{\theta}\left(\mathbf{X}^{0} \mid \mathbf{X}^{1}\right)}_{L_{0}}],
\label{eq:ELBO}
\end{align}
where $D_{KL}$ is the KL divergence between each forward process step and its reconstructed representation in the backwards process.

\subsection{A binomial forward process}
We use a binomial (single trial Bernoulli) Markov diffusion process to fit the binomial input data (see assumptions in Section~\ref{subsec:assumptions}). Intuitively, this corresponds to performing bit flips over diffusion time steps in the forward process and predicting these bit flips in the reverse process. The latter is defined with
\begin{equation}
p_{\theta}\left(\mathbf{X}^{t-1} \mid \mathbf{X}^{t}\right):= \mathcal{B}\left(\mathbf{X}^{t-1} ; \mathbf{\pi}_{\theta}\left(\mathbf{X}^{t}, t\right)\right),
\end{equation}
where $\mathcal{B}(u ; \pi)$ is the distribution for a single Bernoulli trial (bit flip), with $u=1$ occurring with probability $\pi$, and $u=0$ occurring with probability $1-\pi$.
The forward process is a flip of the original $\{0, 1\}$ bits with increasing chance, determined by the schedule $\beta_t$:
\begin{equation}
    q\left(\mathbf{X}^{t} \mid \mathbf{X}^{t-1}\right):=\mathcal{B}\left(\mathbf{X}^{t} ; \mathbf{X}^{t-1}\left(1-\beta_t\right)+0.5 \beta_t\right).
\end{equation}
Let $\alpha_t = 1-\beta_t$ and $\bar{\alpha_t} = \prod_{\tau=1}^t \alpha_\tau$, we get (see Appendix):
\begin{align}
q\left(\mathbf{X}^t|\mathbf{X}^0\right) &= \mathcal{B}\left(\mathbf{X}^{t} ; \bar{\alpha}_t \mathbf{X}^{0}+\frac{1}{2}\left(1-\bar{\alpha}_t\right) \bar{\alpha}_t\right).
\end{align}
This means that, when we use a binomial forward kernel the reverse diffusion process will also be binomial. Then, we can simply use a network to predict the bit flip probability for the reverse process to be modeled accurately.

For the loss function we use a Binary Cross Entropy (BCE) ELBO formulation (Eq.~\ref{eq:ELBO}) for $L_{0}$:
\begin{equation}
L_{0} = -\log p_{\theta}\left(\mathbf{X}^{0} \mid \mathbf{X}^{1}\right) = 
\mathbf{X}^{1} \odot \log \mathbf{X}^{0} + \left(1 - \mathbf{X}^{1}\right) \odot \log \left(1 - \mathbf{X}^{0}\right).
\label{eq:BCEELBO}
\end{equation}
We use $\odot$ as the sign for element-wise multiplication. The traditional ELBO loss relies on the KL divergence. We demonstrate that the KL divergence is also suited for binary data (see Appendix~\ref{appendix:ELBO}). Additionally, we derive the Bernoulli Markov process and verify that a combination (multiplication) of Bernoulli distributions is still a Bernoulli distribution (see Appendix~\ref{appendix:bern}). This guarantees that we can use the Gaussian Markov process properties.

\subsection{RecFusion – Recommendation systems as diffusion models}
\label{subsec:recdiff}
In the image domain, $\mathbf{X}^0$ is of dimension corresponding to the image resolution.\footnote{See Section~\ref{subsec:diff} for our choice of notation.} Instead, in the recommendation setup, $\mathbf{X}^0$ is the full user-item matrix. 

\textbf{Full-matrix.} In our recommendation setting, we can consider the entire user-item matrix as $\mathbf{X}^0$ of dimension $U \times I$, where $U$ is the number of users and $I$ is the number of items. Each cell in that matrix is a binary representation of the feedback of a user on an item (e.g., for MovieLens \cite{movieLens}, $x_{ui}$ is $1$ for ratings above 3 stars and $0$ otherwise, following~\cite{liang2018:multvae}). We are thus framing our setting as non-sequential recommendation with binary feedback (see Section~\ref{sec:exp_setup})).
With ever-growing user-item matrices, it quickly becomes infeasible to perform in-memory computations.
The solution for image diffusion models is to use a first diffusion model for a say $32 \times 32$ image and then use several super-resolution models to upscale it \cite{imagen}. We consider that a user-item matrix cannot be downscaled by blurring it, because it does not contain hierarchical features, unlike for an image (e.g. an image of a dog probably contains the dog's head, which contains eyes, etc.).

\sloppy \textbf{User-batch.} Instead, we could think of batches of users $\mathbf{X}^0_{u \in b_j} \phantom{l} \forall b_j\in\mathbf{b}.$\footnote{Batching by items is also possible, but would rather fit the domain of item-based collaborative filtering \cite{itemBased}.} In that case, the input matrix is still big. For example on MovieLens1M, a batch size $200$ users leads to a $200 \times 10000$ matrix compared to a $32 \times 32$ image, but possible to fit in memory.  There are two more advantages to feeding by batch. (I) We can now perform gradient descent over several examples of the data instead of just one matrix example, and (II) we can form batches of items of the same category and use that as a downstream task (a.k.a. diffusion guidance \cite{ho2021classifierfree}). For example, we could batch by movie genre in the case of the MovieLens dataset~\cite{movieLens}. This user-batch formulation is still similar to the original 2 dimensional image setting, but assumes relationships between users close together in the matrix if convolution-based models are used. This assumption is unrealistic and we thus think that this is not a viable approach theoretically. We nonetheless verify that assumption empirically with \emph{RecFusionU-Net2D}.

\textbf{User-by-user.} Alternatively, we can use a 1D formulation (batch size of 1), with $\mathbf{x}^0_u$, the vector of all item feedbacks for user $u$. In that case, we assume no relationships between users. This corresponds to the formulation of MultVAE~\cite{liang2018:multvae}. With this formulation, the advantages of the user-batch formulation are kept and spatial dependence between users does not need to be assumed. This setting applies to \emph{RecFusion, RecFusionT, RecFusionVar, RecFusionBin, RecFusionU-Net1D}.

We use the vector notation $\mathbf{x}$ for the rest of the paper, to refer to $\mathbf{x}_u$, a user vector. Below we look at two practicalities, conditional generation and the fully perturbed recommendation matrix.

\subsubsection{Generate from $\mathbf{x}^1$, a simple alternative to conditional generation}
In practice, a recommendation platform is interested in finding the top $K$ next items for users (see Assumption 4 in Section~\ref{subsec:assumptions}). In a traditional diffusion inference setup, we would start with a completely random recommendation matrix $\mathbf{x}^{T}$ and generate $\mathbf{x}^0$ iteratively via the backward diffusion pass through the neural network $p_{\theta}\left(\mathbf{x}^{t-1} \mid \mathbf{x}^{t}\right)$ over each diffusion time step $t$. Without any conditioning / guidance / inpainting techniques, the generated matrix $\mathbf{x}^0$ remains the same, given a particular random $\mathbf{x}^T$. We propose a simpler approach: at inference time, we feed the validation/test recommendation matrix as $\mathbf{x}^1$ and perform a single backward diffusion step to $\mathbf{x}^0$.
One question remains: what is a completely perturbed matrix?

\subsubsection{What is a completely perturbed matrix $\mathbf{X}^{T}$ in the recommendation setting?}
\label{sec:perturbed}

Strictly speaking, the kernel $\kappa_{p}$ in the Benoulli markov chain forward process $q\left(\mathbf{X}^{T} \mid \mathbf{X}^{t-1}\right) = \kappa_{p}\left(\mathbf{X}^t|\mathbf{X}^{t-1}; \beta_t\right)$ determines the final state $p(\mathbf{X}^{T})$. Instead, as an experiment, we propose here to start from a desired final state $p(\mathbf{X}^{T})$ and determine a markov chain that leads to it.

\begin{figure*}
\captionsetup[subfigure]{justification=centering}
    \centering
    \begin{subfigure}{0.32\linewidth}
        \centering
        \includegraphics[clip,trim=5mm 8mm 90mm 15mm, width=\linewidth]{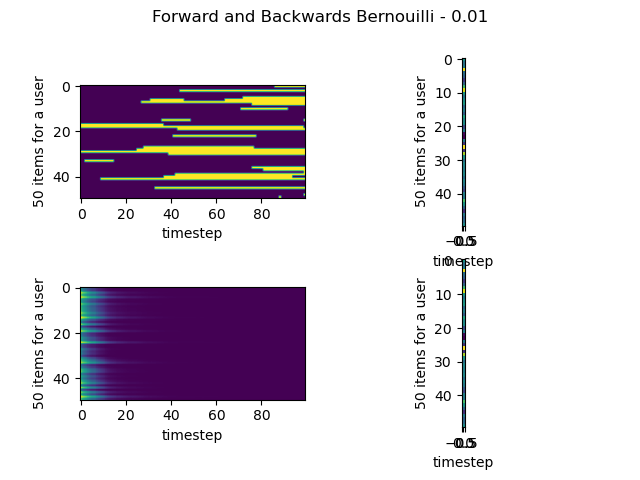}
        \caption{a user vector with same mean\\ activity as the input data $p(x^T = 1) = E(x^0) = \mathbf{\Bar{X}}^0$}
        \label{fig:bindiff0.01}
    \end{subfigure}    %
    \begin{subfigure}[b]{0.32\linewidth}
        \centering
        \includegraphics[clip,trim=5mm 8mm 90mm 15mm, width=\linewidth]{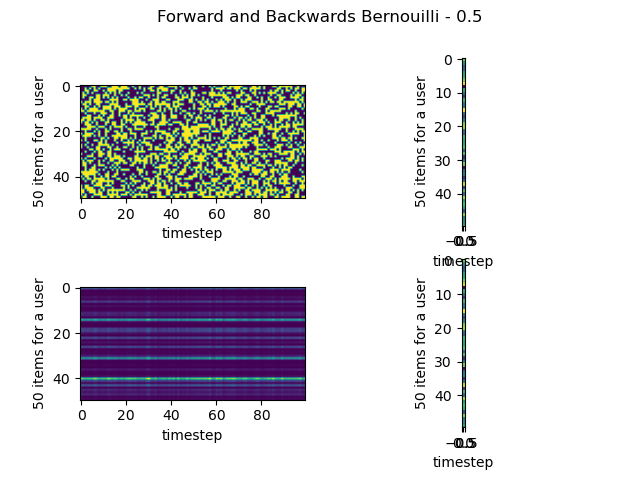}
        \caption{a user vector with a coin flip activity  $p(x^T = 1) = 0.5$ \\ \phantom{XXX} }
        \label{fig:bindiff0.5}
    \end{subfigure} %
    \begin{subfigure}[b]{0.32\linewidth}
        \centering
        \includegraphics[clip,trim=5mm 8mm 90mm 15mm, width=\linewidth]{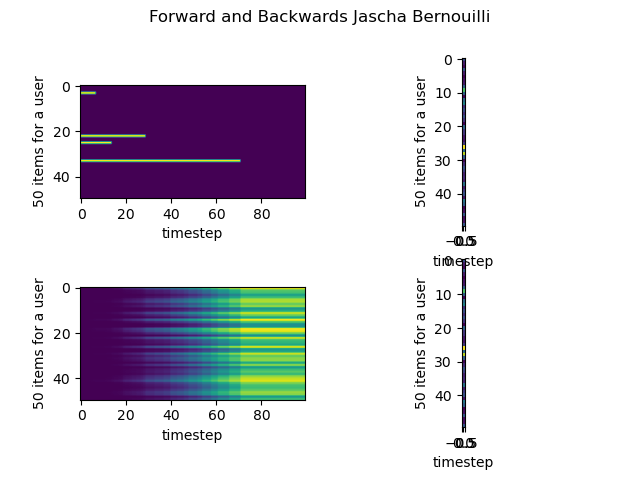}
        \caption{a user vector full of zero values $p(x^T = 1) = 0$ \\ \phantom{XXX} }
        \label{fig:bindiffJascha}
    \end{subfigure}
    \vspace{.5\baselineskip}
    \caption{Binomial diffusion on MovieLens100k after 20 epochs. Top row is the Bernouilli forward process and the bottom row is the learned reverse process. Blue is closer to $0$; yellow is closer to $1$.}
    \label{fig:diffbin}
\end{figure*}

More concretely, we ask what is a completely diffused matrix $\mathbf{X}^T$? Is it (a) a matrix with the same mean activity as the input data $p(\mathbf{X}^T = 1) = E(x^0) = \mathbf{\Bar{X}}^0$ (as proposed by \citeauthor{diffusion} ~\cite{diffusion}) (b) a matrix with a fair coin flip activity  $p(\mathbf{X}^T = 1) = 0.5$ -- in the binomial case $\mathcal{B}(\mathbf{X}^T ; 0.5)$ -- or (c) a matrix full of zero values $p(\mathbf{X}^T = 1) = 0$. We show these three alternatives in Figure~\ref{fig:diffbin} with a Bernouilli diffusion example.

With (a) and (b), we experiment with allowing bit flips from $0$ to $1$ and from $1$ to $0$, by formulating $p_\theta = \mathcal{B}\left(x^{t} ; \beta_t\right)$ and $x^t =  (1 - p_\theta) \cdot x^{t-1} + p_\theta \cdot (1 - x^{t-1})$.
For (a), we use a fixed schedule of $ \beta_t = 0.01$ $\forall t$. The reverse diffusion process is able to pick up a signal.  For (b), we use a fixed schedule of $ \beta_t = 0.5$ $\forall t$. Right away, the user vector becomes chaotic and no real signal is picked up by the reverse diffusion process.
With (c), we only allow bit flips from $1$ to $0$ and end up with with a matrix full of zeroes. For that we let $p_\theta = \mathcal{B}\left(\mathbf{X}^{t} ; \mathbf{X}^{t-1}\left(1-\beta_t\right)+0.5 \beta_t\right)$ and $x^t = p_\theta \cdot x^t$. Again, the reverse diffusion process picks up a signal. We found (c) to work best in practice. We conjecture that this is because bit flips only go in one directions and that this information flows more smoothly in the gradient descent steps.

\subsubsection{Architecture}
We propose a few different architectures for RecFusion, in order of complexity. \emph{RecFusion}, a three layer fully connected network with tanh activation. \emph{RecFusionT} with a time step embedding: we first tried to use the time embedding as in the original attention paper~\cite{attention}, namely feeding the time embedding to the output of the MLP $f(x) + Z_t$. This was not very successful. Instead we fed the time embedding in a DDPM~\cite{DDPM} manner (we are not sure if this practice emerged in DDPM or before): $f(x + Z_t)$. \emph{RecFusionVar}, which predicts mean and variance/error of diffusion steps like in DDPM~\cite{DDPM}. \emph{RecFusionBin} our own 1D Bernoulli diffusion model: forward steps as described in Section~\ref{sec:perturbed} (c), reverse steps with a \emph{RecFusion} architecture but a sigmoid final activation and our own BCE ELBO loss (Eqn.~\ref{eq:ELBO} and~\ref{eq:BCEELBO}). \emph{RecFusionU-Net1D} is the original DDPM~\cite{DDPM} architecture simplified, with only one channel and flattened on one dimension to allow for user vector input $\mathbf{x}$. \emph{RecFusionU-Net2D} is the original DDPM~\cite{DDPM} architecture simplified and only one channel to allow for a user-batch matrix input $\mathbf{X}^0_{u \in b_j} \phantom{l} \forall b_j\in\mathbf{b}$. Both U-Nets have a time embedding.

Our two Unet architectures are more as proof-of-concepts than theoretically grounded architectures. Some elements of the U-Net architecture make it rather impractical, such as the necessary spatial relationships in the matrix/vector and the necessity for an even-sized matrix/vector input for the up-downsizing steps in the Unet. For some datasets, we removed the least popular item from the data altogether, in order to be able to fit an even number of items as a vector / matrix.

%% file: sections/03_experimental_setup.tex
\section{Experimental Setup}
\label{sec:exp_setup}
Our experimental setup focuses on the classical recommendation task, where the task is to predict items which a user would enjoy / interact with, based on historical interactions~\cite{SteckEvaluation}. For instance, prior models like the MultVAE \cite{liang2018:multvae} are fed the user history, and tasked to rank items, where each dimension of the input and output correspond to an item -- in the case of the MultVAE, the predicted likelihood can be used to rank recommended items.

Given original binary input (feedback of whether or not someone liked / consumed an item), it is a bit harder to argue for a regular forward diffusion process with Gaussian noise. %
The forward process is either Gaussian or Binomial.

\subsection{Assumptions}
\label{subsec:assumptions}
Our experiments make the following set of standard assumptions, following prior work: we assume a Top-$K$ recommendation setup for binary implicit feedback, and evaluate using a strong generalization split. These, and other assumptions are explicitly described below:
\begin{description}
\item[Assumption 1] Top-$K$ recommendation: We consider the Top-$K$ recommendation problem, reflected primarily in the evaluation metrics we utilize: Recall@20, Recall@50 and NDCG@100.
\item[Assumption 2] Binary feedback: If a rating is non-binary, we binarize it. We experiment with two datasets: for MovieLens \cite{movieLens} and Netflix \cite{bennett2007:netflix}, we convert ratings of 4 and higher to $1$, and use $0$ otherwise, following prior work \cite{liang2018:multvae, RecVAE, ma2019:macridvae}. 
\item[Assumption 3] Missing or negative interactions cannot be easily distinguished.
Note that this is assumption is typically flawed, but widepsread in the research literature~\cite{Verstrepen2017}. Existing schemes to deal with this typically either 
\begin{enumerate*}[label=(\roman*)]
    \item adopt heuristics to classify user-item interactions as either missing or negative, and weight such instances appropriately in the loss function~\cite{Pan2008,Hu2008}, or
    \item leverage information about users' exposure to recommendations to model this probabilistically~\cite{Liang2016}.
\end{enumerate*}
We envision that such methods can be straightforwardly extended to general classes of diffusion models as well, providing an interesting avenue for future work.
\item[Assumption 4] In contrast to sequential recommendation \cite{wang2019:survey_sequential}, we do not explicitly consider the order in which items are viewed, an assumption consistent with prior work \cite{liang2018:multvae, RecVAE, ma2019:macridvae}, which RecFusion builds on. For the validation and test sets splits, we randomly sample items independently of item consumption time. 
\item[Assumption 5] We filter out users with fewer than five items, and items with fewer than five interactions, as is common practice~\cite{liang2018:multvae,Beel2019e}.
\item[Assumption 6] Strong generalization \cite{marlin2004:cf_ml_perspective}: Users are split into train/validation and test sets, with the training employing the entire history. For validation and test sets, a partial history is fed to the recommender, with a held-out set being used to evaluate the resulting recommendation. 
\end{description}

\subsection{Baselines}
\label{subsec:baselines}
We benchmark our methods against the following \emph{non-neural} baselines:
\begin{enumerate*}[label=(\roman*)]
\item \textbf{Random}: Recommendations are generated by uniformly sampling without replacement from the set of items that have been interacted with.

\item \textbf{Popularity}: The frequency of each item is calculated and subsequently normalized by dividing the individual count by the maximum count among all items. Consequently, every user receives identical recommendations with scores ranging from zero to one.

\item \textbf{SLIM}: Linear model with a sparse item-to-item similarity matrix; solved using a constrained $\ell_1, \ell_2$ regularized optimization problem~\cite{SLIM}.

\item \textbf{EASE}: A variant of SLIM with a closed-form solution, obtained by dropping the non-negativity and $\ell_1$ constraint, which simplifies to ridge regression~\cite{EASE}. 
\end{enumerate*}

We also consider the following \emph{neural} baselines:
\begin{enumerate*}[label=(\roman*)]
\item \textbf{MultVAE}: Variational autoencoder \cite{kingma2013:vae} with a multinomial likelihood~\cite{liang2018:multvae}. 

\item \textbf{RecVAE}: Improves upon the MultVAE with a composite prior, newer architecture and a training schedule which alternates between training the encoder/decoder~\cite{RecVAE}.  

\item \textbf{CODIGEM}: We took the original CODIGEM code and had to to fix some bugs to make it run. Once it ran in the original bare repo, we transferred the modelling code to the RecPack framework~\cite{walker-2022-recommendation}.
\end{enumerate*}

\subsection{Implementation and parameters}
We provide a model card in the Appendix~\ref{appendix:modelCard}. We use RecPack \cite{michiels_verachtert2022:recpack}, a reproducibility framework for our experiments. We reproduce baselines ourselves, given the ambiguity over the aforementioned assumptions in existing literature. We promote a reproducible setup  with the above assumptions. 

We utilize the following datasets in our experiments 
\begin{enumerate*}[label=(\roman*)]
\item MovieLens \cite{movieLens} (we use 1M, 25M)
\item Netflix \cite{bennett2007:netflix}
\end{enumerate*}. Dataset statistics are reported in Appendix~\ref{appendix:descStats}. As mentioned before, we evaluate on the test set using the following metrics: Recall@20, Recall@50 and NDCG@100. We report \textit{calibrated recall}, which adjusts for the number of true positive interactions and ensure that optimal recommendations map to a perfect recall value of 1, as is commonly done in previous work~\cite{liang2018:multvae}. We train on single NVIDIA V100 GPUs.

For hyperparameter tuning, we use Hyperopt~\cite{hyperopt} and its Tree of Parzen Estimators~\cite{tpe} algorithm and Sparktrial\footnote{\url{http://hyperopt.github.io/hyperopt/scaleout/spark/}} to coordinate GPUs. We use the validation set NDCG@100 to navigate the hyperparameter space. Once the best hyperparameter combination is found, we run the model on the test split (train/val/test – $0.8/0.1/0.1$). For MovieLens1M, we bootstrap predictions and run on 10 different splits to obtain error bars on out-of-sample prediction (see Figure~\ref{fig:ML1M}).

%% file: sections/04_results.tex
\section{Results}
Our results show that between the diffusion models, RecFusion outperforms CODIGEM on two of three datasets. However non-neural baselines, and EASE in particular, outperform both neural and diffusion-based models on all datasets.

\begin{figure*}
    \centering
    \includegraphics{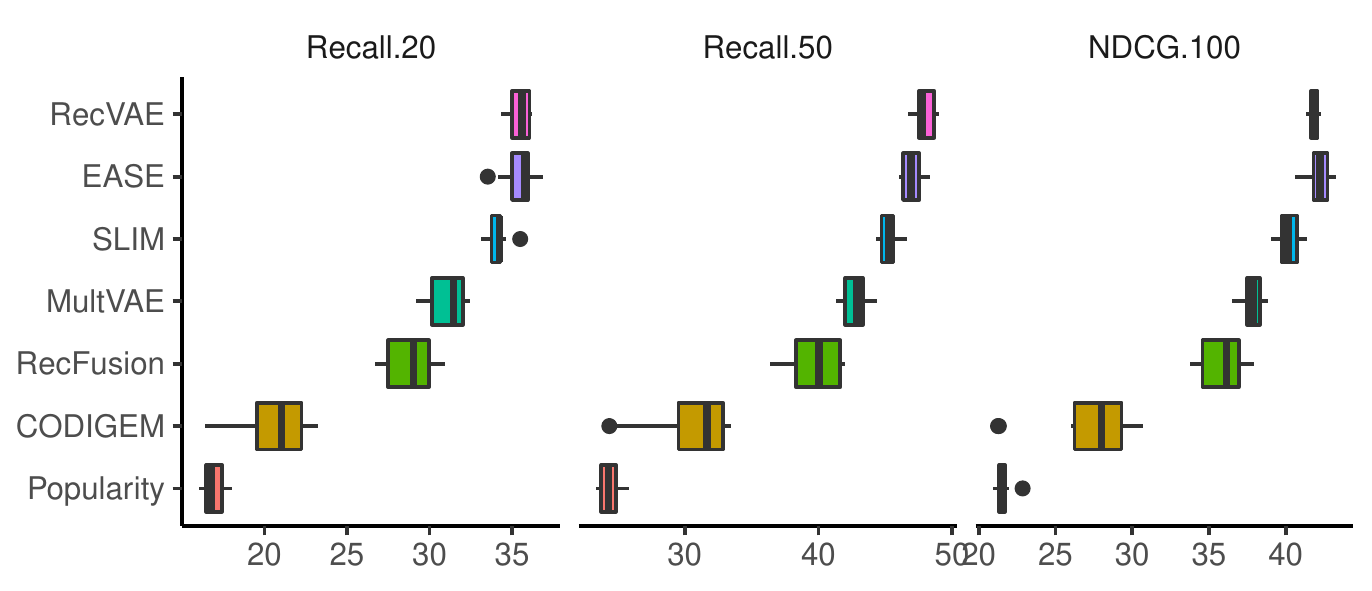}
    \caption{Experimental results on the MovieLens1M dataset. All results reproduced by us. Our method is RecFusion.  Boxplots show median and IQR over 10 train/test splits}.
    \label{fig:ML1M}
\end{figure*}

\textbf{Non-neural baselines.} %
Across datasets and metrics, the best performance is often obtained by EASE (see Table \ref{fig:ML1M} and Table \ref{tab:gen}). EASE even outperforms MultVAE, a popular neural baseline, on most datasets and metrics. This is in line with prior research that demonstrates the efficacy of linear models for recommendation over some neural methods \cite{EASE, dacrema2019:repro_study, dacrema2021:repro_followup}. Despite this, neural methods make other tasks within recommendation viable (e.g., using user or item metadata), as highlighted in Section~\ref{sec:relatedWork}. %

\begin{table*}[b]
\centering
\caption{Experimental results on the MovieLens25M and Netflix datasets. All results are reproduced by us. Our method in bold.}
\begin{tabular}{llrrr}
\toprule
Dataset & Model & Recall@20 &  Recall@50 &  NDCG@100   \\
\midrule
\multirow{6}{*}{MovieLens25M} & Random &       0.13 &       0.30 &      0.24 \\
        & Popularity &      16.63 &      24.43 &     19.69 \\
        & \textbf{RecFusion} &      33.21 &      45.44 &     37.31 \\
        & CODIGEM &      34.05 &      45.84 &     37.90 \\
        & MultVAE &      35.12 &      48.09 &     39.12 \\
        & EASE &      40.02 &      52.71 &     43.84 \\
\midrule        
\multirow{6}{*}{Netflix} & Random &       0.18 &       0.32 &      0.31 \\
        & Popularity &      11.73 &      17.48 &     15.89 \\
        & CODIGEM &      25.54 &      33.48 &     29.08 \\
        & \textbf{RecFusion} &      29.68 &      37.63 &     32.87 \\
        & MultVAE &      31.61 &      40.61 &     35.23 \\
        & EASE &      36.19 &      44.49 &     39.35 \\
\bottomrule
\end{tabular}
\label{tab:gen}
\end{table*}

\textbf{Comparing diffusion models and neural methods.} From Figure \ref{fig:ML1M} and Table \ref{tab:gen}, we observe a consistent trend: MultVAE outperforms both diffusion models, CODIGEM and RecFusion, across all datasets and for all metrics. %
One reason might be the difference in the number of parameters employed by the two networks: MultVAE uses two three-layer networks, one each for the encoder and decoder, whereas RecFusion employs a single three-layer network, which is reflected in the number of parameters reported in Table~\ref{tab:model_nparams}. For MoveLens 1M, we observe that RecVAE %
outperforms MultVAE and both diffusion models. However, RecVAE uses a somewhat complicated (per-user) prior, along with a complicated training schedule where only the encoder (or decoder) is trained with the decoder (or encoder) is frozen. RecFusion employs none of these heuristics. 

We highlight that RecFusion outperforms, or is on par with CODIGEM. We keep this time the most popular model in each of VAEs and non-neural classifications in Table~\ref{tab:ML1M}.

\textbf{Ablation study.}  In Table~\ref{tab:ablation}, we perform an ablation study: we start with models that integrate the most diffusion methods and remove elements, one-by-one. Perhaps unsurprisingly for recommendations where linear models dominate, we discover that the most bare-bone (close to linear) diffusion model works best. RecFusionU-Net1D and RecFusionU-Net2D drastically under-perform, even scoring below the Popularity baselines. For RecFusionUNet-2D, this is expected because of the lack of spatial correlations that the model was originally designed for. 

Adding typical diffusion elements like time embeddings (RecFusionT), mean/variance (RecFusionVar) also underperforms compared to the base RecFusion model. We hypothesize that RecFusionBin underperforms due to the noise schedule employed: adding noise to images (256 colors) is more meaningful than adding noise to binary data. We exacerbate this problem by explicitly modelling it as a binomial Markov diffusion process.

\begin{table*}[t]
\vspace{-.5\baselineskip}
\centering
\caption{Experimental results for different RecFusion methods on MovieLens1M.}
\begin{tabular}{lrrr}
\toprule
Model &   Recall@20 &  Recall@50 &  NDCG@100\\
\midrule
RecFusionU-Net1D &       4.45 &       7.77 &      6.99\\
RecFusionU-Net2D &       6.47 &       9.08 &      9.03\\
RecFusionT &      14.03 &      17.80 &     16.59\\
RecFusionVar &      16.71 &      24.73 &     21.63\\
RecFusionBin &      17.59 &      23.53 &     21.94\\
RecFusion &      30.91 &      41.76 &     37.44\\
\bottomrule
\end{tabular}
\label{tab:ablation}
\vspace{-.5\baselineskip}
\end{table*}

\textbf{Summary of results and discussion.}
Our results show that existing VAE (MultVAE, RecVAE) and non-neural baselines (EASE, SLIM) outperform more complicated architectures, like diffusion models. RecFusion, however, outperforms the diffusion baseline, CODIGEM, on two of three datasets across all metrics. RecFusion is the simplest form of our approach, with a VAE akin to MultVAE (in terms of mean-variance estimation and loss function). We stress that our proposed method, RecFusion is a simpler and more elegant way to model the recommendation problem than CODIGEM. Our experiments, however, highlight the difficulty of utilizing generative models for real-world problems in which (close to) linear models dominate.

%% file: sections/05_related_work.tex
\section{Related Work}
\label{sec:relatedWork}
This work should not be confused with diffusion models in social recommendation (e.g., \cite{socialDiffusion1, socialDiffusion2}), an orthogonal field. We briefly review the diffusion and the recommmendation literature.

\textbf{Diffusion models.}
Diffusion probabilistic models were first introduced by~\cite{diffusion}, where the specific implementation and optimization objectives failed to surpass the state-of-the-art. A few years later, the denoising diffusion model (DDPM) was introduced by~\cite{DDPM}, where the loss function is simplified and the architecture proposed manages to achieve strong state-of-the-art performance. The rich literature that follows would be impossible to summarize in a single paragraph. The most relevant work is denoising diffusion implicit models which changes the parametrization of the sampling to make it deterministic instead of stochastic \citep{songdenoising}. Diffusion is first often used in the 2D domain, it can also have a 3D interpretation \citep{ho-2022-video, chen23:singl_stage_diffus_nerf, yang23:learn_diffus_prior_nerfs, diff_3D_medical}.

Diffusion for recommendation is already present in early work. CODIGEM is probably the first attempt at using diffusion models for recommendation~\cite{walker-2022-recommendation}. They take inspiration from diffusion models and generate recommendations through iterative denoising. Although Diffusion models inspire CODIGEM, it is implemented effectively as a simple hierarchical variational autoencoder. The first reason is that the model does not share weights across timesteps. Also, diffusion models are based on the assumption that the forward process is performed in sufficiently small steps to guarantee that the reverse will have the same functional form \cite{diffusion,Feller1949OnTT}.

\textbf{Recommender systems.}
Non-neural MF methods can solve minimization problems on single user-item rating matrices (see Figure~\ref{fig:one}). But 
\begin{enumerate*}[label=(\roman*)]
\item user-item metadata,
\item time representation,
\item and controllability / guidance (e.g., a movie recommendation set that must be action-comedy oriented)
\end{enumerate*}
are harder to model in a closed form or iterative manner (e.g., Gibbs sampler for ALS \cite{menzen-2021-alternating}). This is where neural models can help. Within neural models, probabilistic models and especially Variational Auto Encoders (VAEs) are omnipresent, including MultVAE~\cite{liang2018:multvae} and RecVAE~\cite{RecVAE}.

Recommendation (together with, arguably, time series and tabular data) is one of only few areas where neural models do not seem to have gained supremacy yet. This has been shown in the settings of general recommendation~\cite{dacrema2019:repro_study, nonNeuralClassic2, nonNeuralClassic3, nonNeuralClassic4}, sparse interactions~\cite{nonNeuralSparse}, session-based~\cite{nonNeuralSessionBased} and next basket recommendation~\cite{nonNeuralBasket1, nonNeuralBasket2, li-2021-next-arxiv}. In these benchmarks, winning methods are variations of matrix factorization (MF) techniques (SVD++, (i)ALS, EASE~\cite{EASE}, and SLIM~\cite{SLIM}) or even the \emph{most popular} benchmark.
Neural models are a popular choice for recommender systems, with early models like AutoRec \cite{suvash2015:autorec} or CDAE \cite{wu2016:cdae} employing auto-encoder architectures. Despite limited reproducibility of some neural models, \cite{dacrema2019:repro_study, dacrema2021:repro_followup,li-2021-next-arxiv} or the superior performance of non-neural methods in certain settings \cite{EASE, SLIM}, e.g., competitions \cite{nonNeuralClassic2}, neural methods have comparable or better performance in several settings. Of these, probabilistic methods employing \textit{variational} inference, i.e., variational auto-encoders (VAE) \cite{kingma2013:vae}, like the MultVAE \cite{liang2018:multvae} or RecVAE \cite{RecVAE} are notable, with the latter being the only neural model successfully reproduced in a large-scale reproducibility study \cite{dacrema2019:repro_study, dacrema2021:repro_followup}.  

\textbf{Contemporaneous work.}
In April 2023, while the table of results above was being finalized, three papers were published on diffusion for recommendation and one on Bernoulli diffusion (some peer-reviewed). \emph{BSPM}~\cite{blur} uses score-based models as a testbed for generative models for the recommendation. \emph{DiffuRec}~\cite{seqdiffrec} is the first attempt at diffusion for sequential recommendation and thus not comparable to our non-sequential setup. \emph{DiffRec} is a similar paper to ours on smaller datasets~\cite{diffRec}. \emph{DiffRec} corresponds to one of our Recfusion formulations (RecFusionVar)  but results on ML1M are significantly (3X) lower compared to our computations or~\cite{sachdeva2022infinite}. This highlights the difficulty of comparisons in the recommendation literature, due to different experimental setups and (at times, unstated) data pre-processing assumptions. We make these assumptions explicit in our paper and code (see Section~\ref{subsec:assumptions}). Similarly, it was a challenge to bring the CODIGEM code to work in general (the code does not run as-is) and within our framework in particular (see Section \ref{subsec:baselines}). However, we come fairly close to the original numbers reported in the paper. Finally, \emph{BerDiff}~\cite{berdiff} is the first attempt we could find to explicitly model binary data with a Bernoulli Markov diffusion process.\footnote{\citet{diffusion} already hinted at diffusion as a 1-dimensional idea as a proof of concept. We could not find the full mathematical derivation or code for it, however.} \emph{BerDiff} focuses on 2D CT scan and MRI data and thus relies heavily on the Unet architecture. In our paper, we show theoretically and empirically that we face more of a 1D problem and thus define our own 1D diffusion model for binary data.

%% file: sections/06_discussion.tex
\section{Conclusion}
\vspace{-.3\baselineskip}

We position this paper as a first attempt at designing diffusion models for unstructured binary 1D data in the context of recommendation and beyond. With RecFusion, our simple diffusion model (hierarchical VAE) is on par with popular VAE methods. We conjecture that extensions (techniques like composite priors, etc as in RecVAE \cite{RecVAE}) can further improve performance. We first argue that we need to tackle limitations in our existing implementation and lay out some proposals for future improvements.
We can summarize our contributions as follows. First, we show theoretically and empirically that the lack of spatial relationship between users and items is a hindrance to using any image-inspired models, including even a 1D U-Net. We then implemented our binary (Bernoulli) Markov process, as a model adapted to the problem at hand.

\textbf{Broader impact.}
The image domain sometimes still requires the simplicity of binary settings, like segmentation masks on MRI, CT scans~\cite{berdiff, ma2023segment} or for conventional object detection techniques~\cite{kirillov2023segment}. 
This is potentially fruitful ground for applying our proposed diffusion models for binary 1D data.

\section{Limitations}
\vspace{-.3\baselineskip}

Our setup relies on weak (even if common) recommendation setup assumptions. To these assumptions we have to add that the items list is fixed: our model can not account for new items in the catalogue after training. This is a limitation shared with CODIGEM, but also with VAE-based models. 

We have yet to test how robust our diffusion models are to a relaxations of these assumptions. Diffusion can also be applied to further domains of recommendation like sequential recommendation with diffusion + RNN, or explicitly model count data input with star ratings instead of binarized feedback.

RecFusion does not yet use further diffusion methods, such as inpainting, guidance (e.g., to predict the user preference distribution or use a prior on movie genre a.k.a controllable recommendations), diffusion on the embedding space~\cite{gao2023difformer} (in particular, user-item matrix embeddings), or multinomial likelihood to model the dependencies of item feedbacks for a user~\cite{multinomialDiffusion}, input masking.
We believe these are fruitful areas for future work.

\section*{Acknowledgments}

We are grateful to Lien Michiels who co-developed the recommendation framework RecPack and answered many questions about how to bend RecPack to our needs; Christian Andersson Naesseth who helped in discussions about inpainting; Nathalie de Jong for designing the first version of Figure\ref{fig:one}. This research was (partially) funded by Bertelsmann SE \& Co. KGaA, by the Hybrid Intelligence Center, a 10-year program funded by the Dutch Ministry of Education, Culture and Science through the Netherlands Organisation for Scientific Research, \url{https://hybrid-intelligence-centre.nl}, and project LESSEN with project number NWA.1389.20.183 of the research program NWA ORC 2020/21, which is (partly) financed by the Dutch Research Council (NWO).

All content represents the opinion of the authors, which is not necessarily shared or endorsed by their respective employers and/or sponsors.

%% file: appendix.tex
\newpage

\appendix

\section{The ELBO is also suited for Bernoulli samples}
\label{appendix:ELBO}

According to the classic definition of the ELBO by~\cite{diffusion} and~\cite{DDPM}, there are no assumptions regarding the distributions $p_\theta$ or $q_\theta$. We reproduce here for completeness the derivation from~\cite{diffusion} on why the ELBO satisfies any distribution given Jensen's inequality:
\begin{equation*}
\begin{aligned}
L & =\int d \mathbf{X}^0 q\left(\mathbf{X}^0\right) \log p\left(\mathbf{X}^0\right) \\
& =\int d \mathbf{X}^0 q\left(\mathbf{X}^0\right) \log 
\left[
\begin{array}{c}
\int d \mathbf{X}^{1 :  T} q\left(\mathbf{X}^{1 :  T} \mid \mathbf{X}^0\right) \\
p\left(\mathbf{X}^{T}\right) \prod_{t=1}^T \frac{p\left(\mathbf{X}^{t-1} \mid \mathbf{X}^{T}\right)}{q\left(\mathbf{X}^{T} \mid \mathbf{X}^{t-1}\right)}
\end{array}
\right].
\end{aligned}
\end{equation*}
The latter has a lower bound given Jensen's inequality that also applies to the bernoulli distribution.
\begin{equation*}
\begin{aligned}
L \geq & \int d \mathbf{X}^{0 :  T} q\left(\mathbf{X}^{0 :  T}\right) \log \left[p\left(\mathbf{X}^{T}\right) \prod_{t=1}^T \frac{p\left(\mathbf{X}^{t-1} \mid \mathbf{X}^{T}\right)}{q\left(\mathbf{X}^{T} \mid \mathbf{X}^{t-1}\right)}\right].
\end{aligned}
\end{equation*}
In practice, the product term is computed with a KL divergence. It can be shown with Fano's inequality~\cite{scarlett_cevher_2021} that our cross-entropy loss also aims for a lower bound like KL divergence.

For this assumption regarding $p_\theta$ or $q_\theta$ to be valid we make sure that the forward steps (i.e. $\beta_t$) are small enough, following~\cite{diffusion}.

\section{Proof of why Bernoulli diffusion is multiplicative}
\label{appendix:bern}

Given our Bernoulli diffusion formulation
\begin{equation*}
    q\left(\mathbf{X}^{t} \mid \mathbf{X}^{t-1}\right):=\mathcal{B}\left(\mathbf{X}^{t} ; \mathbf{X}^{t-1}\left(1-\beta_t\right)+{\textstyle \frac{1}{2}} \beta_t\right),
\end{equation*}
we would like to make sure that $q\left(\mathbf{X}^{t} \mid \mathbf{X}^{t-1}\right)$ can still be sampled at an arbitrary step $t$ in closed form, as with traditional gaussian diffusion~\cite{diffusion}. Without loss of generalization -- since we sample independently from a Bernoulli distribution -- we show that this is true for a single user-item combination. Let $x^t$ be the random variable that represents the $t$-th forward diffusion step for a specific user-item combination. Then:
\begin{equation*}
\begin{aligned}
x^t & =\left(1-\beta_t\right) x^{t-1}+{\textstyle \frac{1}{2}} \beta_t.
\end{aligned}
\end{equation*}
By substituting $x^{t-1}$, we get
\begin{equation*}
\begin{aligned}
x^t & =\left(1-\beta_t\right)\left[\left(1-\beta_{t-1}\right) x_{t-2}+{\textstyle \frac{1}{2}} \beta_{t-1}\right]+{\textstyle \frac{1}{2}} \beta_t \\
& =\left(1-\beta_t\right)\left(1-\beta_{t-1}\right) x_{t-2}+{\textstyle \frac{1}{2}}\left(1-\beta_t\right) \beta_{t-1}+ {\textstyle \frac{1}{2}} \beta_t.
\end{aligned}
\end{equation*}
If we keep on substituting the previous diffusion step, we arrive at the original data $x_0$. By factorizing the above and by induction, it is trivial to show that 
\begin{equation}
\label{eq:beforeTelescope}
\begin{aligned}
x^t & = \prod_{i=1}^t\left(1-\beta_i\right) x_0+\frac{1}{2} \sum_{j=1}^{t-1}\left[\prod_{i=j+1}^t\left(1-\beta_i\right) \right] \beta_j  + \frac{1}{2} \beta_t.
\end{aligned}
\end{equation}
We can actually express the middle term with a common index by using $\beta_j = 1- (1 - \beta_j)$.\footnote{We borrow this trick from \url{https://math.stackexchange.com/questions/4467894/does-a-markov-chain-with-gaussian-transitions-px-tx-t-1-mathcal-n-sqrt1}} We then obtain a telescoping sum:
\begin{equation*}
\begin{aligned}
\sum_{j=1}^{t-1}\left[\prod_{i=j+1}^t\left(1-\beta_i\right) \right] \beta_j & =  \sum_{j=1}^{t-1}\left[\prod_{i=j+1}^t \left(1-\beta_i\right)\left(1-\left(1-\beta_j\right)\right) \right]\\
& = \sum_{j=1}^t \left[ \prod_{i=j+1}^t\left(1-\beta_i\right)-\prod_{i=j}^t\left(1-\beta_i\right) \right] = 1-\prod_{i=1}^t\left(1-\beta_i\right).
\end{aligned}
\end{equation*}
Substituting this term back into Equation \ref{eq:beforeTelescope},
\begin{equation*}
\begin{aligned}
x^t & = \prod_{i=1}^t\left(1-\beta_i\right) x_0+ {\textstyle \frac{1}{2}} \left[ 1-\prod_{i=1}^t\left(1-\beta_i\right) \right] + {\textstyle \frac{1}{2}} \beta_t.
\end{aligned}
\end{equation*}
Finally, by defining $\alpha_t = 1-\beta_t$ and $\bar{\alpha_t} = \prod_{\tau=1}^t \alpha_\tau$, we get
\begin{align}
x^t &=\bar{\alpha}_t x_0+\frac{1}{2}\left(1-\bar{\alpha}_t\right) \bar{\alpha}_t \\
q\left(\mathbf{X}^t|\mathbf{X}^0\right) &= \mathcal{B}\left(\mathbf{X}^{t} ; \bar{\alpha}_t \mathbf{X}^{0}+\frac{1}{2}\left(1-\bar{\alpha}_t\right) \bar{\alpha}_t\right).
\end{align}
We showed that $\mathbf{X}^t$ can be sampled directly from $\mathbf{X}^0$ in a single bernoulli sample. \hfill $\blacksquare$

\section{Model card}
\label{appendix:modelCard}

See \url{https://github.com/gabriben/recfusion/blob/master/model_card.md}.

\section{Descriptive statistics}
\label{appendix:descStats}

In Table \ref{tab:data-stats}, we show counts of users items and interactions on the train, val and test sets. We provide this as an extra step for data preprocessing transparency. 

\begin{table}[h]
\centering
\caption{Descriptive statistics: Counts of active (non-zero) users, items and interactions after preprocessing and under train / val / test (0.8 / 0.1 / 0.1) splitting regime over users.}
\begin{adjustbox}{width=\textwidth, center}
\begin{tabular}{@{}lrrrrrrrrr@{}}
\toprule
       & \multicolumn{3}{c}{No. users} & \multicolumn{3}{c}{No. items} & \multicolumn{3}{c}{No. interactions} \\
       \cmidrule{2-4} \cmidrule(lr){5-7} \cmidrule{8-10}
Dataset       & train & val & test & train & val & test & train & val & test \\ \midrule
ML1M  &    4,832 & 604 & 604   &     3,416 & 3,158 & 3,282 &   798,608 & 76,513 & 84,772 \\
ML25M &      130,032 & 16,254 & 16,255 &     32,718 & 24,818 & 25,597      &   19,924,515 & 1,999,297 & 2,030,221 \\
Netflix       &        378,389 & 47,299 & 47,299 &  17,769 & 17,761 & 17,759         & 
 80,418,808 & 8,011,940 & 8,060,214 \\
\bottomrule
\end{tabular}%
\end{adjustbox}
\label{tab:data-stats}
\end{table}

\section{Results on MovieLens1M as a table}
\label{appendix:ML1M}

Table~\ref{tab:ML1M} is the pendant of Figure \ref{fig:ML1M} for the MovieLens 1M dataset. We added the Random baseline here, but left it out of the figure for aesthetics.

\begin{table*}[th!]
\centering
\caption{Experimental results on the MovieLens1M dataset. All results reproduced by us over 10 train/test splits, we report median results. Our method in bold.}
\label{tab:ML1M}
\begin{tabular}{llrrr}
\toprule
type & model      & Recall@20 &  Recall@50 &  NDCG@100\\
\midrule
\multirow{2}{*}{baselines} & Random     &       0.87 &       1.71 &      1.73 \\
 & Popularity &      16.77 &      24.30 &     21.44 \\
\midrule
\multirow{2}{*}{diffusion} & CODIGEM    &      21.04 &      31.67 &     28.00 \\
& \textbf{RecFusion}  &      29.02 &      40.03 &     36.14 \\
\midrule
\multirow{2}{*}{VAE} & MultVAE    &      31.43 &      42.89 &     37.87 \\
& RecVAE     &      35.61 &      47.79 &     41.81 \\
\midrule
\multirow{2}{*}{non-neural} & SLIM       &      34.23 &      45.25 &     40.14 \\
& EASE       &      35.76 &      46.92 &     42.24 \\
\bottomrule
\end{tabular}
\vspace*{-3mm}
\end{table*}

\newpage

\section{Number of parameters}
\label{appendix:numParam}

One of our arguments is that our model is more efficient than existing neural baselines (see Table \ref{tab:model_nparams}).

\begin{table}[h]
    \centering
    \caption{Number of parameters for different neural architectures on the ML1M dataset.}
    \begin{tabular}{ccc}
    \toprule
    MultVAE  & CODIGEM & RecFusion\\
    \midrule
    446,421,6  & 560,388,6 & 141,021,8 \\
    \bottomrule         
    \end{tabular}
    
    \label{tab:model_nparams}
\end{table}